\documentclass[aps,prl,showpacs,footinbib,reprint,superscriptaddress,twocolumn,longbibliography]{revtex4-1}
\usepackage{amsmath,amssymb,amsfonts,mathtools,bm,dcolumn,mathrsfs,color,braket,graphicx}
\usepackage[colorlinks=true,linkcolor=blue,citecolor=blue,urlcolor=blue]{hyperref}

\begin{document}

\title{Boundary-Controlled Liouvillian Relaxation with Exact Steady States Fixed by Dissipative Disorder}

\author{Y. Z. Miao}
\affiliation{School of Science, Qingdao University of Technology, Qingdao, Shandong, China}
\author{W. Z. Ma}
\affiliation{School of Science, Qingdao University of Technology, Qingdao, Shandong, China}
\author{Y. Wang}
\affiliation{School of Science, Qingdao University of Technology, Qingdao, Shandong, China}
\author{X. L. Zhao}
\email{zhaoxiaolong@qut.edu.cn}
\affiliation{School of Science, Qingdao University of Technology, Qingdao, Shandong, China}
\author{X. X. Yi}
\email{yixx@nenu.edu.cn}
\affiliation{Center for Quantum Sciences and School of Physics, Northeast Normal University, Changchun, Jilin, China}

\begin{abstract}
In open quantum lattice systems, changing the boundary condition would appear to alter both
the steady state and the nonzero Liouvillian spectrum. Here we show that boundary conditions
can be used to control relaxation without changing the reduced steady state. In a disordered
dissipative quantum link chain, the steady state is determined by an accumulated
field defined by link-resolved dissipative disorder, and a gauge-generated transformation
built from this field gives exact symmetry-resolved steady states with nonuniform,
accumulated-field-dependent reduced matter occupations. We then construct a reciprocal cyclic
boundary condition that preserves these matter occupations while changing the nonzero
Liouvillian spectrum. Consequently, open and cyclic chains relax to the same reduced matter
steady-occupation profile with different Liouvillian gaps with the cyclic closure accelerating
relaxation. In the strong-dissipation limit, this relaxation difference can be reduced to a spectral
comparison of effective exclusion processes with open and cyclic boundaries.
\end{abstract}

\maketitle

\emph{Introduction.---}
Boundary sensitivity is well known in discussions of non-Hermitian skin effects, where
changing the boundary condition can reshape the spectrum and eigenmode
profiles~\cite{Bergholtz2021,Okuma2023,Ashida2020,Zhang2022Review,
Yao2018,Lee2019,Kunst2018,Song2019Chiral,Ghatak2020,Weidemann2020,
Xiao2020,Helbig2020}. In open quantum lattice systems, the same boundary sensitivity
directly affects Lindbladian relaxation: the boundary condition can modify both the steady
state and the nonzero Liouvillian modes that govern the relaxation toward it
~\cite{Haga2021,Mao2024Liouvillian,Feng2024Boundary}. When these two changes
occur together, the observed relaxation cannot be assigned uniquely to a
changed steady state or to a changed nonzero spectrum.

Removing this ambiguity requires a steady observable that remains fixed when
the boundary condition is changed. Globally reciprocal non-Hermitian systems show how such
boundary-insensitive profiles can arise when local nonreciprocal biases with
vanishing global bias produce realization-dependent profiles controlled by an
accumulated field~\cite{Longhi2025Erratic,Longhi2026Liouvillian,Miao2026AAH}.
Extending this accumulated-field control to interacting open quantum systems is
nontrivial. Related many-body skin phenomena have been studied in systems with
dynamical gauge couplings~\cite{Faugno2022,Li2023Gauge}, exactly solvable
interacting nonreciprocal dynamics~\cite{Zheng2024ExactBH}, and boundary-sensitive
Liouvillian chains~\cite{Mao2024Liouvillian,Wang2023ScaleFree}. However, for disordered
interacting dissipative chains, exact steady states and boundary-resolved
nonzero Liouvillian spectra are generally unavailable. It therefore remains
unclear whether a disordered accumulated field can fix the occupation of matter at
steady state while the boundary condition changes the nonzero Liouvillian spectrum.

Quantum link chains provide the candidate for addressing this
question. Link-local dissipation defines local dissipative asymmetries, while
constrained matter-link exchange relates the corresponding link
weights to matter-site weights. This structure follows from gauge constraints,
which couple matter and link degrees of freedom and restrict the allowed
dynamics~\cite{Kogut1979,Kogut1983,Fradkin2013}, and is naturally formulated
within finite-dimensional quantum link models~\cite{Banerjee2012atomic}.
Relevant gauge dynamics has been implemented or proposed in ultracold atoms
~\cite{Zohar2015,Mil2020,Yang2020,Zhou2022}, trapped ions
~\cite{Hauke2013,Martinez2016}, Rydberg arrays~\cite{Cheng2024}, and
superconducting circuits~\cite{Marcos2013,Blais2021}, with parallel progress
in gauge protection~\cite{Stannigel2014Constrained,Halimeh2020Reliability,
Halimeh2020Robustness,Halimeh2021Protection}, controlled dissipation
~\cite{Syassen2008,Tomita2019}, and dissipative gauge steady states
~\cite{Wang2024TopologicalOrdered,Dai2025SteadyTopological}. In the uniformly
biased case, dissipative quantum link models admit exact many-body skin steady
states~\cite{Hu2025}, motivating the dissipative quantum
link chain with globally reciprocal link-dissipation disorder studied here.

Here we focus on a chain with locally asymmetric but globally
reciprocal link dissipation. We show that the disordered dissipative asymmetries
define an accumulated field, from which a gauge-generated transformation
gives exact symmetry-resolved steady states under open boundary conditions
(OBC) and accumulated-field ordered reduced matter steady occupations. We then
construct a reciprocal cyclic boundary condition (CBC), which preserves the OBC reduced matter steady
occupation but changes the nonzero Liouvillian spectrum, thereby modifying the
Liouvillian gap and the late-time matter-occupation relaxation rate. This separates the accumulated-field control of the steady occupation from the boundary control of relaxation, and the separation admits a concise spectral interpretation in terms of effective open and cyclic exclusion processes in the strong-dissipation limit.

\emph{Model and symmetries.---}
We first consider a one-dimensional $U(1)$ quantum link chain with $L$ matter sites and OBC ~\cite{Banerjee2012atomic,Chandrasekharan1997,HuangDQPT2019}.
In the fermionic form, the coherent Hamiltonian is
$H_{\rm f}=J\sum_{n=1}^{L-1}
(\psi_n^\dagger U_{n,n+1}\psi_{n+1}+\mathrm{H.c.})$,
where $\psi_n$ acts on matter site $n$, and $U_{n,n+1}$ is conjugate to the
electric field $E_{n,n+1}$, satisfying
$[E_{n,n+1},U_{n,n+1}]=U_{n,n+1}$ and
$[E_{n,n+1},U_{n,n+1}^{\dagger}]=-U_{n,n+1}^{\dagger}$. With the links oriented
from left to right, the lattice divergence of the
electric field at a bulk site is $E_{n,n+1}-E_{n-1,n}$. The fermionic
Gauss-law generators are therefore
$\mathcal G_n^{\rm f}=\psi_n^\dagger\psi_n-(E_{n,n+1}-E_{n-1,n})$ for
$1<n<L$, with boundary forms
$\mathcal G_1^{\rm f}=\psi_1^\dagger\psi_1-E_{1,2}$ and
$\mathcal G_L^{\rm f}=\psi_L^\dagger\psi_L+E_{L-1,L}$. For fixed static
background charges $g_n$, the Gauss-law sector is the simultaneous eigenspace
of the local generators satisfying
$\mathcal G_n^{\rm f}\ket{\Phi}=-g_n\ket{\Phi}$ for all sites $n$.

After representing the quantum links by spin operators and
applying the Jordan--Wigner transformation to the matter fields~\cite{Hu2025,Chandrasekharan1997,HuangDQPT2019},
the Hamiltonian $H_{\rm f}$ becomes
\begin{equation}\label{eq:hamiltonian}
H
=
J\sum_{n=1}^{L-1}
\left(
\tau_n^+s_{n,n+1}^+\tau_{n+1}^-
+\mathrm{H.c.}
\right).
\end{equation}
In this correlated exchange form, $\tau_n^{\pm}$ act on matter site $n$, while
$s_{n,n+1}^{\pm}$ act on the intervening link $(n,n+1)$, with
$s_{n,n+1}^z$ representing the electric field and $[s_{n,n+1}^z,s_{m,m+1}^{\pm}]=
\pm\delta_{nm}s_{m,m+1}^{\pm}$. We take the minimal link spins $s=1/2$ throughout. The matter occupation and total
matter number are $\hat n_n=\tau_n^+\tau_n^-=\tau_n^z+1/2$ and
$\hat N=\sum_{n=1}^{L}\hat n_n$, so that $\tau_n^z=1/2$ ($-1/2$) denotes an
occupied (empty) site.
After absorbing the uniform matter offset into the static background charges,
the spin Gauss generators are
\begin{equation}\label{eq:gauge_generators}
G_n
=
\tau_n^z-\bigl(s_{n,n+1}^z-s_{n-1,n}^z\bigr),
\qquad
1<n<L.
\end{equation}
For the two open boundaries, we have
$G_1=\tau_1^z-s_{1,2}^z$ and
$G_L=\tau_L^z+s_{L-1,L}^z$. The spin Gauss generators satisfy $[G_n,H]=0$ and
$\sum_{n=1}^{L}G_n=\sum_{n=1}^{L}\tau_n^z=\hat N-L/2$.

\begin{figure}[t]
\centering
\includegraphics[width=\linewidth]{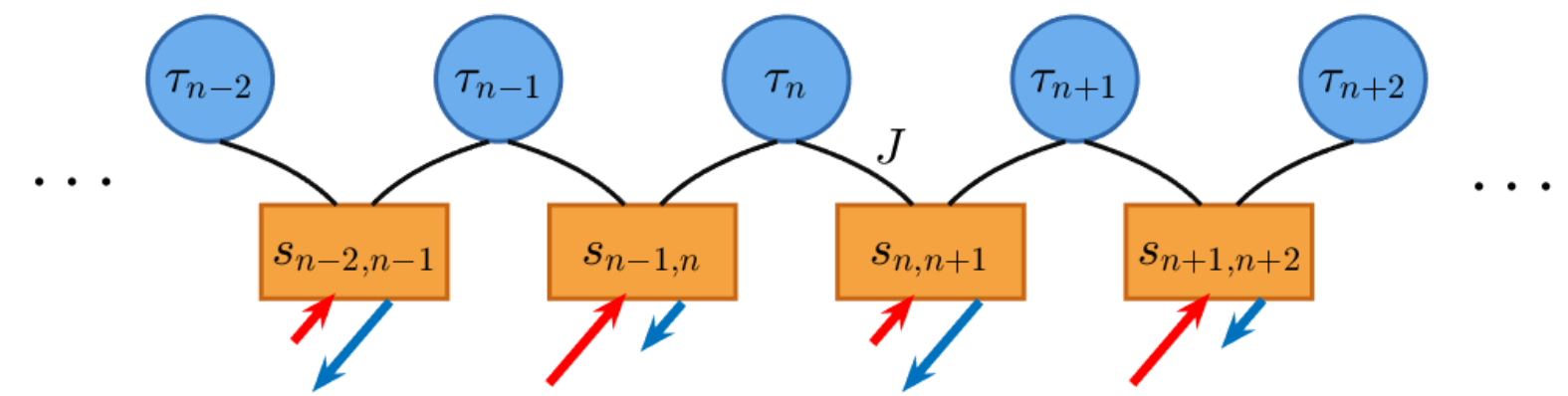}
\caption{
Dissipative $U(1)$ quantum link chain. Blue circles denote matter spins
$\tau_n$, orange rectangles denote link spins $s_{n,n+1}$, black curves denote
the coherent coupling $J$, and red/blue arrows denote link pumping and loss.
}
\label{fig:model}
\end{figure}

The coherent dynamics described by Eq.~\eqref{eq:hamiltonian} is a
gauge-invariant correlated exchange between matter and link degrees of freedom,
while the dissipation is taken to act directly on the links, as illustrated in
Fig.~\ref{fig:model}. The open-system dynamics is governed by
\begin{equation}\label{eq:lindblad}
\begin{aligned}
\frac{d\rho}{dt}
&=
\mathcal L[\rho]
=
-i[H,\rho] \\
&\quad
+
\sum_{n=1}^{L-1}\sum_{\mu=u,d}
\left(
2L_n^{(\mu)}\rho L_n^{(\mu)\dagger}
-
\{L_n^{(\mu)\dagger}L_n^{(\mu)},\rho\}
\right).
\end{aligned}
\end{equation}
We decompose $\mathcal L=\mathcal L_H+\mathcal L_D$, with
$\mathcal L_H[\rho]\equiv -i[H,\rho]$ and $\mathcal L_D$ denoting the dissipative
part in Eq.~\eqref{eq:lindblad},
in which
$L_n^{(u)}=\sqrt{\gamma_{u,n}}\,s_{n,n+1}^{+}$ and
$L_n^{(d)}=\sqrt{\gamma_{d,n}}\,s_{n,n+1}^{-}$. The local dissipative asymmetry
on link $n$ is denoted by
$h_n=\ln(\gamma_{u,n}/\gamma_{d,n})$. Equivalently,
$\gamma_{u,n}=\gamma e^{h_n/2}$ and
$\gamma_{d,n}=\gamma e^{-h_n/2}$, with $\gamma>0$ setting the overall
dissipation scale. We consider the globally reciprocal case
\begin{equation}\label{eq:global_reciprocal}
\sum_{n=1}^{L-1}h_n=0,
\end{equation}
so that the total dissipative asymmetry vanishes while finite local disordered dissipative
asymmetries on the links are retained.

The jump operators break the strong gauge symmetry generated by $G_n$, but
preserve a weak Liouvillian symmetry. Defining the weak-gauge superoperator
$\mathcal W_n[\rho]\equiv G_n\rho-\rho G_n$, one has
$[\mathcal W_n,\mathcal L]=0$. The matter number remains a strong symmetry
because $[\hat N,H]=0$ and $[\hat N,L_n^{(\mu)}]=0$. The Liouvillian therefore
decomposes in operator space into blocks labeled by the eigenvalues of
$\mathcal W_n$ and by the conserved matter number. The steady states and spectra
considered below are evaluated in the zero weak-gauge block
$\mathcal W_n[\rho]=0$ at fixed matter number $N$.

\emph{Accumulated-field steady state.---}
We construct the exact OBC steady state by extending the link-space
transformation of the dissipator to a gauge-generated transformation on the
matter-link chain. Since $\mathcal L_D$ acts independently on different links,
its steady state in the link sector is diagonal and factorizes as
\begin{equation}\label{eq:rho_s_main}
\rho_s
=
\bigotimes_{n=1}^{L-1}\rho_{n,n+1},
\qquad
\rho_{n,n+1}
=
\frac{
e^{h_n}\ket{\uparrow}\bra{\uparrow}
+
\ket{\downarrow}\bra{\downarrow}
}{
1+e^{h_n}
},
\end{equation}
where $\ket{\uparrow}$ and $\ket{\downarrow}$ are the $\pm1/2$ eigenstates of
$s_{n,n+1}^z$. The diagonal weights in Eq.~\eqref{eq:rho_s_main} determine the
diagonal link-space transformation
$T_s=\bigotimes_{n=1}^{L-1}T_{n,n+1}$, with
$T_{n,n+1}\propto e^{-h_ns_{n,n+1}^z}$, which maps $\rho_s$ to the identity,
$T_s\rho_s\propto I_s$. With the left-multiplication superoperator
$\mathcal T_s[\rho]=T_s\rho$, applying this link-space transformation to each
local link dissipator gives
$\mathcal T_s\mathcal L_D\mathcal T_s^{-1}=\mathcal L_D^\dagger$.

The link factors $e^{-h_ns_{n,n+1}^z}$ alone would rescale the correlated
exchange terms in Eq.~\eqref{eq:hamiltonian}. To extend the link-space
transformation to a gauge-generated transformation that leaves these terms
invariant, the matter-site weights must satisfy nearest-neighbor differences
equal to the local dissipative asymmetries $h_n$. We therefore introduce the
accumulated field
\begin{equation}\label{eq:accumulated_field}
X_1=0,
\qquad
X_n=\sum_{m=1}^{n-1}h_m,
\qquad n\ge2 ,
\end{equation}
thus $X_{n+1}-X_n=h_n$. We use the gauge-generated transformation
$T=\exp[-\sum_{n=1}^{L}(\ln\alpha+X_n)G_n]$, where $\alpha>0$ is an arbitrary
constant. Using Eq.~\eqref{eq:gauge_generators}, the exponent decomposes as
$\sum_{n=1}^{L}(\ln\alpha+X_n)G_n
=
\sum_{n=1}^{L}(\ln\alpha+X_n)\tau_n^z
+
\sum_{n=1}^{L-1}(X_{n+1}-X_n)s_{n,n+1}^z$.
Since $X_{n+1}-X_n=h_n$, the link part of $T$ differs from $T_s$ only by an
overall scalar factor. Equation~\eqref{eq:global_reciprocal} gives
$X_L=X_1=0$, while the intermediate partial sums in
Eq.~\eqref{eq:accumulated_field} depend on the disorder realization and define
the accumulated-field profile.

Since $\mathcal L_D$ acts only on the links and the link part of $T$ matches
$T_s$ up to an overall scalar factor, the left-multiplication superoperator
$\mathcal T[\rho]=T\rho$ gives
$\mathcal T\mathcal L_D\mathcal T^{-1}=\mathcal L_D^\dagger$.
Since $[G_n,H]=0$ for all $n$, the transformation leaves the Hamiltonian~\eqref{eq:hamiltonian}
invariant, $THT^{-1}=H$. Namely, the factors generated on
$\tau_n^+$, $s_{n,n+1}^+$, and $\tau_{n+1}^-$ cancel inside each correlated
exchange term. Hence
$\mathcal T\mathcal L\mathcal T^{-1}=\mathcal L_H+\mathcal L_D^\dagger$.
Because $\mathcal L_H[I]=0$ and $\mathcal L_D^\dagger[I]=0$, the identity is a
steady operator of the transformed Liouvillian. Transforming back gives the
exact OBC steady state
\begin{equation}\label{eq:exact_steady}
\rho_{\mathrm{ss}}
=
\frac{\mathcal T^{-1}[I]}{\mathrm{Tr}\{\mathcal T^{-1}[I]\}}
=
Z^{-1}
\exp\left[
\sum_{n=1}^{L}(\ln\alpha+X_n)G_n
\right].
\end{equation}
Since $\sum_{n=1}^{L}G_n=\hat N-L/2$, the term
$\ln\alpha\sum_nG_n=\ln\alpha(\hat N-L/2)$ depends only on the conserved matter
number. Thus $\ln\alpha$ can be viewed as the chemical potential of a Gibbs
ensemble. After projection to a fixed eigenvalue $N$ of $\hat N$, $\ln\alpha$ is
a constant and is removed by normalization. The symmetry-resolved OBC steady
state in the fixed-$N$ sector is therefore
\begin{equation}\label{eq:sector_ss}
\rho_{\mathrm{ss},N}
=
Z_N^{-1}\,
P_N
\exp\left[
\sum_{n=1}^{L}X_nG_n
\right]
P_N ,
\end{equation}
where $P_N$ projects onto the eigenspace of $\hat N$ with eigenvalue $N$, and
$Z_N=\mathrm{Tr}\!\{P_N\exp[\sum_{n=1}^{L}X_nG_n]P_N\}$.

The fixed-$N$ reduced matter distribution is obtained by tracing out the link
spins in Eq.~\eqref{eq:sector_ss}. In the exponent
$\sum_{\ell=1}^{L}X_\ell G_\ell$, the link contribution is
$\sum_{q=1}^{L-1}(X_{q+1}-X_q)s_{q,q+1}^z$ and gives a
matter-configuration-independent factor after tracing out the link degrees of freedom. The matter
contribution is $\sum_{\ell=1}^{L}X_\ell\tau_\ell^z$. For a matter
occupation configuration $\mathbf a=(a_1,\ldots,a_L)$, where $a_\ell=0,1$
is the occupation of site $\ell$, one has
$\tau_\ell^z=a_\ell-1/2$. The factors independent of $\mathbf a$ are
absorbed into the normalization, and the fixed-$N$ reduced matter
distribution is
\begin{equation}\label{eq:reduced_matter_distribution}
p_N(\mathbf a)
=
\delta_{(\sum_{\ell=1}^{L}a_\ell,N)}\,
\frac{
\exp\!\left(\sum_{\ell=1}^{L}X_\ell a_\ell\right)
}{
Z_N^{\rm m}
},
\end{equation}
with
$Z_N^{\rm m}
=
\sum_{\mathbf a}
\delta_{(\sum_{\ell=1}^{L}a_\ell,N)}
\exp\!\left(\sum_{\ell=1}^{L}X_\ell a_\ell\right)$.
The site-resolved reduced matter steady occupation is
$N_n^{\mathrm{ss}}\equiv
\mathrm{Tr}[\rho_{\mathrm{ss},N}\hat n_n]=\sum_{\mathbf a}p_N(\mathbf a)a_n$.

\begin{figure}[t]
\centering
\includegraphics[width=\linewidth]{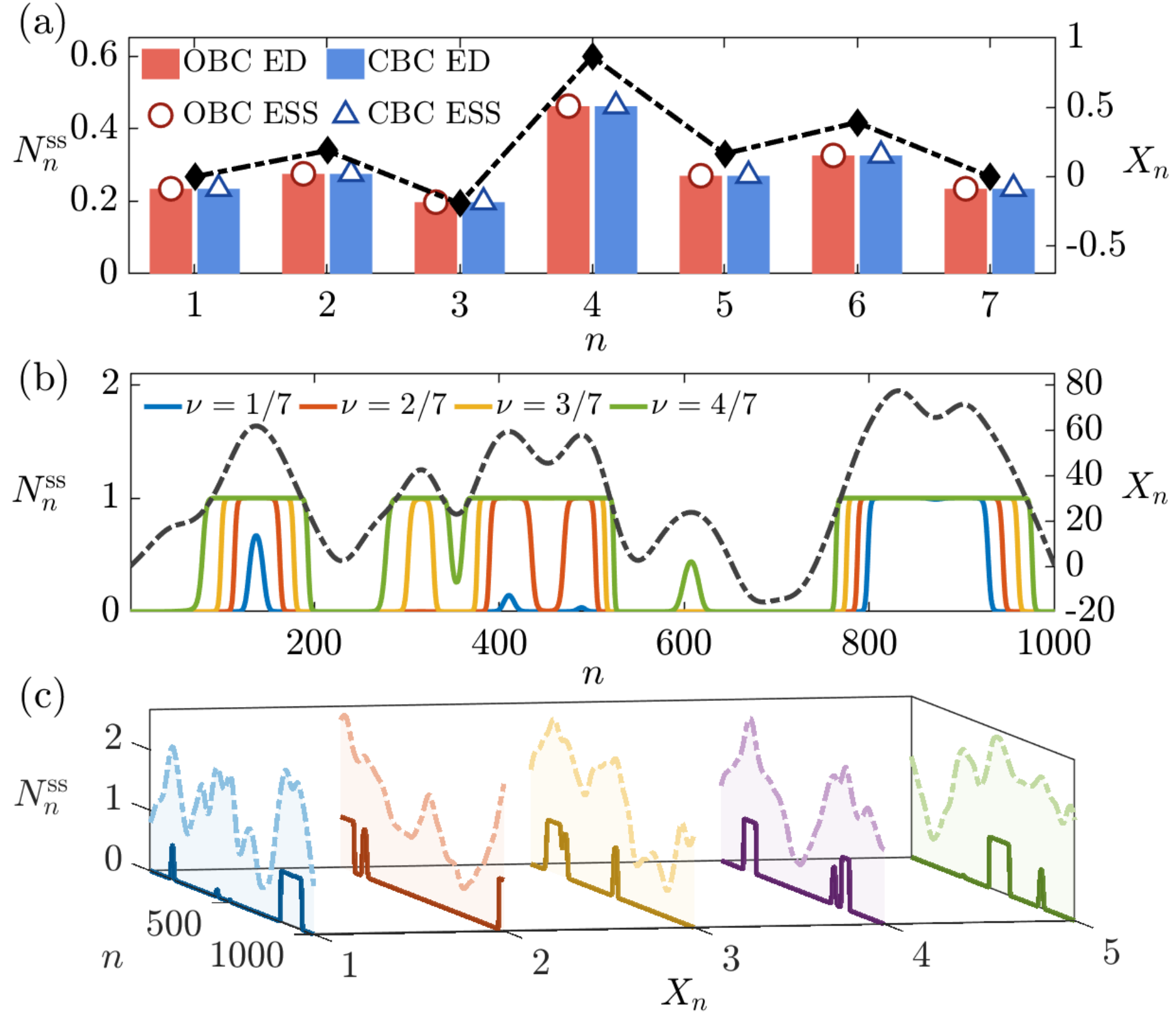}
\caption{
Accumulated-field control of $N_n^{\mathrm{ss}}$.
In all panels, $N_n^{\mathrm{ss}}$ is plotted against the left axis and the
accumulated field $X_n$ against the right axis.
(a) $L=7$, $N=2$: bars show OBC and CBC values of
$N_n^{\mathrm{ss}}$ from exact diagonalization, markers show the corresponding
values from exact symmetry-resolved steady states, and the dash-dotted curve
shows $X_n$.
(b) OBC $N_n^{\mathrm{ss}}$ from the exact reduced matter distribution for a
separate $L=1001$ realization at
$\nu\equiv N/L=1/7,\,2/7,\,3/7,\,4/7$; the dash-dotted curve shows $X_n$ for
the same realization.
(c) OBC $N_n^{\mathrm{ss}}$ from the exact reduced matter distribution for five
realizations with $L=1001$, $N=143$; dash-dotted curves show the corresponding
$X_n$.
Parameters are $J=1$ and $\gamma=2$. For each realization, the interior
asymmetries $h_n$ are constructed from globally reciprocal Gaussian disorder
realizations with exact sample mean zero and sample variance
$\sigma_h^2=0.4$.
}
\label{fig:numerics_main}
\end{figure}

To show how the accumulated field orders the site occupations, compare two
distinct matter sites $i\ne j$ for $0<N<L$. From Eq.~\eqref{eq:reduced_matter_distribution},
$N_i^{\mathrm{ss}}-N_j^{\mathrm{ss}}
=\sum_{\mathbf a}p_N(\mathbf a)(a_i-a_j)$. Only configurations with
$(a_i,a_j)=(1,0)$ or $(0,1)$ contribute to
$N_i^{\mathrm{ss}}-N_j^{\mathrm{ss}}$. For each admissible remaining configuration
$\mathbf b=\{b_\ell\}_{\ell\ne i,j}$ satisfying
$\sum_{\ell\ne i,j}b_\ell=N-1$, the two contributing configurations
$(a_i,a_j)=(1,0)$ and $(0,1)$ have weights proportional to
$e^{X_i}\exp(\sum_{\ell\ne i,j}X_\ell b_\ell)$ and
$e^{X_j}\exp(\sum_{\ell\ne i,j}X_\ell b_\ell)$, respectively.
Subtracting their contributions and summing over all admissible
$\mathbf b$ gives
\begin{equation}\label{eq:density_difference_main}
N_i^{\mathrm{ss}}-N_j^{\mathrm{ss}}
=
\frac{
\left(e^{X_i}-e^{X_j}\right) W^{(i,j)}_{N-1}
}{
Z_N^{\rm m}
},
\end{equation}
with
$W^{(i,j)}_{N-1}
=
\sum_{\mathbf b}
\delta_{(\sum_{\ell\ne i,j}b_\ell,N-1)}
\exp\!\left(\sum_{\ell\ne i,j}X_\ell b_\ell\right)>0$.
Thus
\begin{equation}\label{eq:density_ordering}
X_i>X_j
\quad\Longleftrightarrow\quad
N_i^{\mathrm{ss}}>N_j^{\mathrm{ss}}.
\end{equation}

Equations~\eqref{eq:reduced_matter_distribution} and
\eqref{eq:density_ordering} give the accumulated-field control of the matter
steady-occupation profile. In Fig.~\ref{fig:numerics_main}(a), the OBC
$N_n^{\mathrm{ss}}$ from Eq.~\eqref{eq:reduced_matter_distribution} agrees
with exact diagonalization in the zero weak-gauge block at fixed matter number,
and the profile is ordered by $X_n$. For another realization of $X_n$, Fig.~\ref{fig:numerics_main}(b) shows that changing
the filling factor redistributes the occupation over the same accumulated-field
profile, with the site occupations following Eq.~\eqref{eq:density_ordering}.
Across different globally reciprocal disorder realizations,
Fig.~\ref{fig:numerics_main}(c) shows that different accumulated-field $X_n$
produce different OBC matter steady-occupation profiles, each ordered according
to Eq.~\eqref{eq:density_ordering}. Thus the
many-body erratic steady occupation is controlled by the realization-dependent
accumulated field.

\emph{Reciprocal boundary condition and Liouvillian branches.---}
In boundary-sensitive open systems, changing the boundary generally changes both
the fixed-$N$ reduced matter steady-state distribution and the nonzero
Liouvillian spectrum~\cite{Hu2025,Feng2024Boundary}.
Equation~\eqref{eq:reduced_matter_distribution} shows that the two boundary
responses can be disentangled at the steady-state level: this distribution is
preserved if the matter-site values $\{X_n\}$ are kept unchanged. We therefore
close the chain by imposing cyclic consistency on the accumulated field.

We implement this closure by keeping the $L-1$ interior links fixed and adding
a single link between sites $L$ and $1$. Let $J_b$ denote the coherent
correlated-exchange coupling strength on the closing link, and let $h_L$ denote its
local dissipative asymmetry. Keeping the interior values of $X_n$ fixed, a
single-valued accumulated field on the cycle means $X_{L+1}=X_1$,
equivalently $\sum_{n=1}^{L}h_n=0$. Since the interior realization already
satisfies Eq.~\eqref{eq:global_reciprocal}, the added link must have
$h_L=0, \gamma_{u,L}=\gamma_{d,L}\equiv g_0$, where $g_0>0.$
Throughout this work we set $g_0=\gamma$ and $J_b=J$. Thus, within this
single-link closure scenario, cyclic consistency of the accumulated field
determines the reciprocal dissipative closing link, while the coherent exchange is
closed with the same coupling strength as in the interior.

With this reciprocal closing link, the cyclic spin Gauss generators are
$G_n^{\mathrm{CBC}}=\tau_n^z-(s_{n,n+1}^z-s_{n-1,n}^z)$, with cyclic link
indices, and the accumulated field satisfies $X_{L+1}=X_1$. The CBC steady state
in the fixed matter-number sector is
\begin{equation}\label{eq:sector_ss_C}
\rho_{\mathrm{ss},N}^{\mathrm{CBC}}
=
Z_{\mathrm{CBC},N}^{-1}\,
P_N
\exp\left[
\sum_{n=1}^{L}X_nG_n^{\mathrm{CBC}}
\right]
P_N.
\end{equation}
Tracing out the links gives the same fixed-$N$ reduced matter distribution as
Eq.~\eqref{eq:reduced_matter_distribution}. Hence the reduced matter steady
occupation is unchanged:
\begin{equation}\label{eq:same_reduced_occupation}
N_{n,\mathrm{CBC}}^{\mathrm{ss}}
=
N_n^{\mathrm{ss}},
\qquad
n=1,\ldots,L.
\end{equation}
This equality is shown in Fig.~\ref{fig:numerics_main}(a), where the OBC and
CBC steady occupations coincide.

\begin{figure}[t]
\centering
\includegraphics[width=0.85\linewidth]{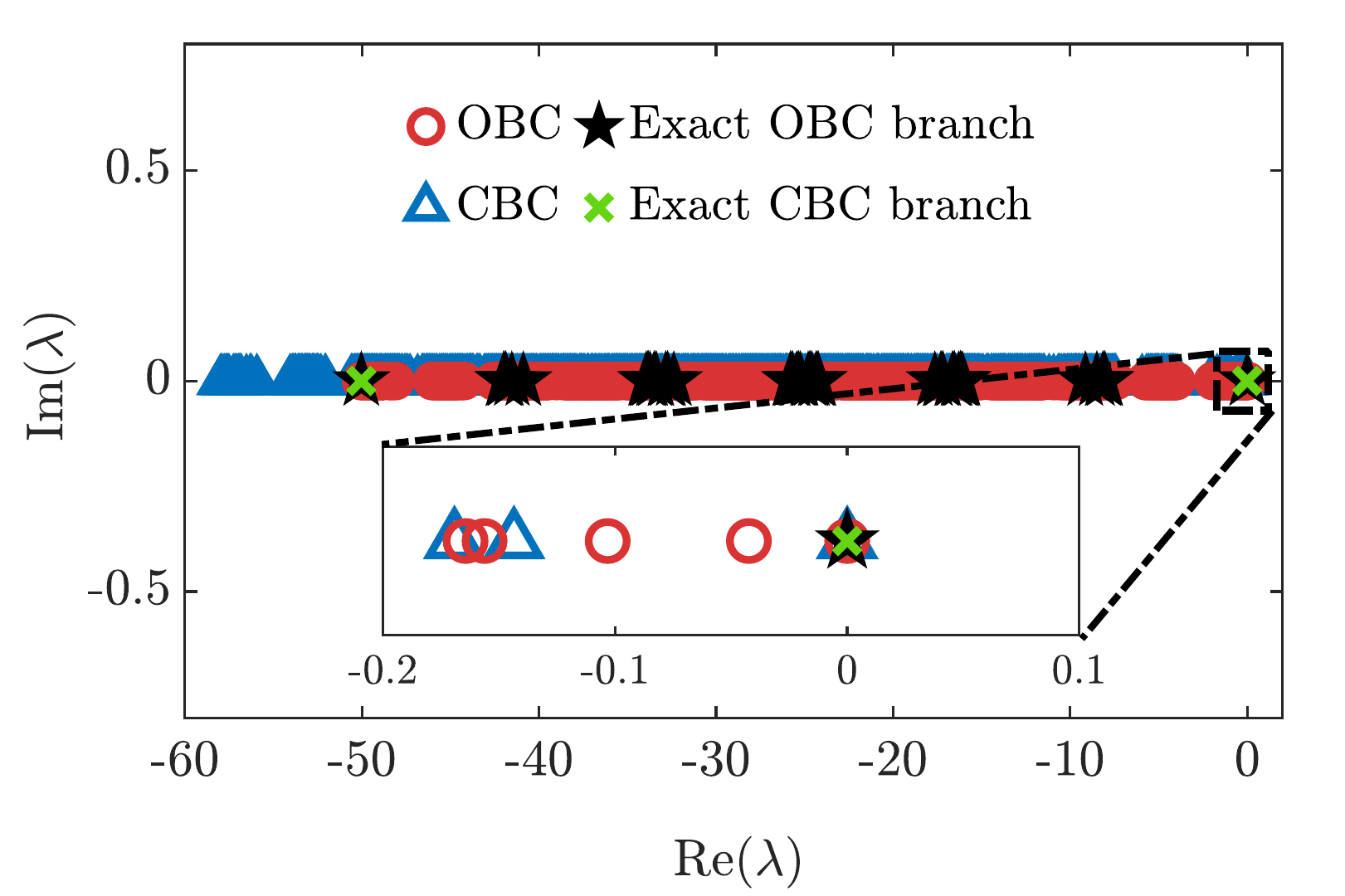}
\caption{
Liouvillian spectra in the zero weak-gauge block at fixed matter number.
Red: OBC. Blue: CBC. Black stars: exact OBC branch eigenvalues from Eq.~\eqref{eq:exact_branch_family}.
Green crosses: CBC branch eigenvalues obtained from the cycle-consistent branch assignments satisfying Eq.~\eqref{eq:pbc_branch_closure}.
The inset enlarges the zero-mode region.
Parameters are the same as in Fig.~\ref{fig:numerics_main}(a).
}
\label{fig:sp}
\end{figure}

The preservation of the reduced matter distribution does not hold in the nonzero
Liouvillian spectrum: the full OBC and CBC spectra in Fig.~\ref{fig:sp}
differ away from the zero mode. We now use the gauge-generated construction to compare
the exact branch eigenoperators determined by the two boundary conditions and thereby identify the
analytic signature for their nonzero spectral difference. Suppose that an operator-space transformation
$\mathcal T_\lambda[\rho]=T_\lambda\rho$ leaves $\mathcal L_H$ invariant and satisfies
$\bigl(\mathcal T_\lambda \mathcal L_D \mathcal T_\lambda^{-1}\bigr)[I]=\lambda I$.
Then $\varrho_\lambda=\mathcal T_\lambda^{-1}[I]$ is a right eigenoperator of
$\mathcal L$ with eigenvalue $\lambda$. For the present link dissipator, each link has two local branches: the zero
branch $x_{0,n}=h_n$, $\lambda_{0,n}=0$, and the nonzero branch
$x_{1,n}=i\pi$, $\lambda_{1,n}=-2(\gamma_{u,n}+\gamma_{d,n})$; see
Supplemental Material~\cite{supplemental}.
For OBC, the branch choice can be made independently on each interior link. Let
$k_n\in\{0,1\}$, $\mathbf{k}=k_1k_2\cdots k_{L-1}$, and define
$X_1^{(\mathbf{k})}=0$ and
$X_n^{(\mathbf{k})}=\sum_{m=1}^{n-1}x_{k_m,m}$ for $n\ge2$. The corresponding
exact OBC eigenoperator is
\begin{equation}\label{eq:exact_branch_family}
\varrho_{\mathbf{k}}
=
\exp\!\left[
\sum_{n=1}^{L}X_n^{(\mathbf{k})}G_n
\right],
\qquad
\lambda_{\mathbf{k}}
=
-2\sum_{n=1}^{L-1}k_n(\gamma_{u,n}+\gamma_{d,n}).
\end{equation}
The assignment $\mathbf{k}=\mathbf{0}$ gives the steady state. Assignments with
at least one $k_n=1$ give exact nonzero OBC eigenoperators, whose eigenvalues are marked by
the black stars in Fig.~\ref{fig:sp}.
For a branch assignment under CBC, the factor
$\exp(\sum_{n=1}^{L}x_{k_n,n})$ must be single-valued after one loop.
Equivalently, $\sum_{n=1}^{L}x_{k_n,n}$ may change only by an integer multiple
of $2\pi i$, which gives the cyclic consistency condition
\begin{equation}\label{eq:pbc_branch_closure}
\exp\left(
\sum_{n=1}^{L}x_{k_n,n}
\right)=1.
\end{equation}
For the zero branch, Eq.~\eqref{eq:pbc_branch_closure} reduces to
$\sum_{n=1}^{L}h_n=0$, which is precisely the accumulated-field consistency
condition that determines the reciprocal CBC. For nonzero branch assignments under
CBC, Eq.~\eqref{eq:pbc_branch_closure} imposes an additional cyclic
consistency condition absent under OBC. The assignments satisfying this
condition give the CBC branch eigenvalues marked by the green crosses in
Fig.~\ref{fig:sp}. The exact OBC branch eigenoperators associated with the
black-star eigenvalues are generally not compatible with the same condition of Eq.~\eqref{eq:pbc_branch_closure} and
have no CBC counterparts. The reciprocal CBC therefore preserves the reduced
matter steady occupation through the zero branch, whereas the nonzero Liouvillian
spectrum remains boundary-sensitive.

\emph{Boundary-controlled relaxation.---}
\begin{figure}[t]
\centering
\includegraphics[width=\linewidth]{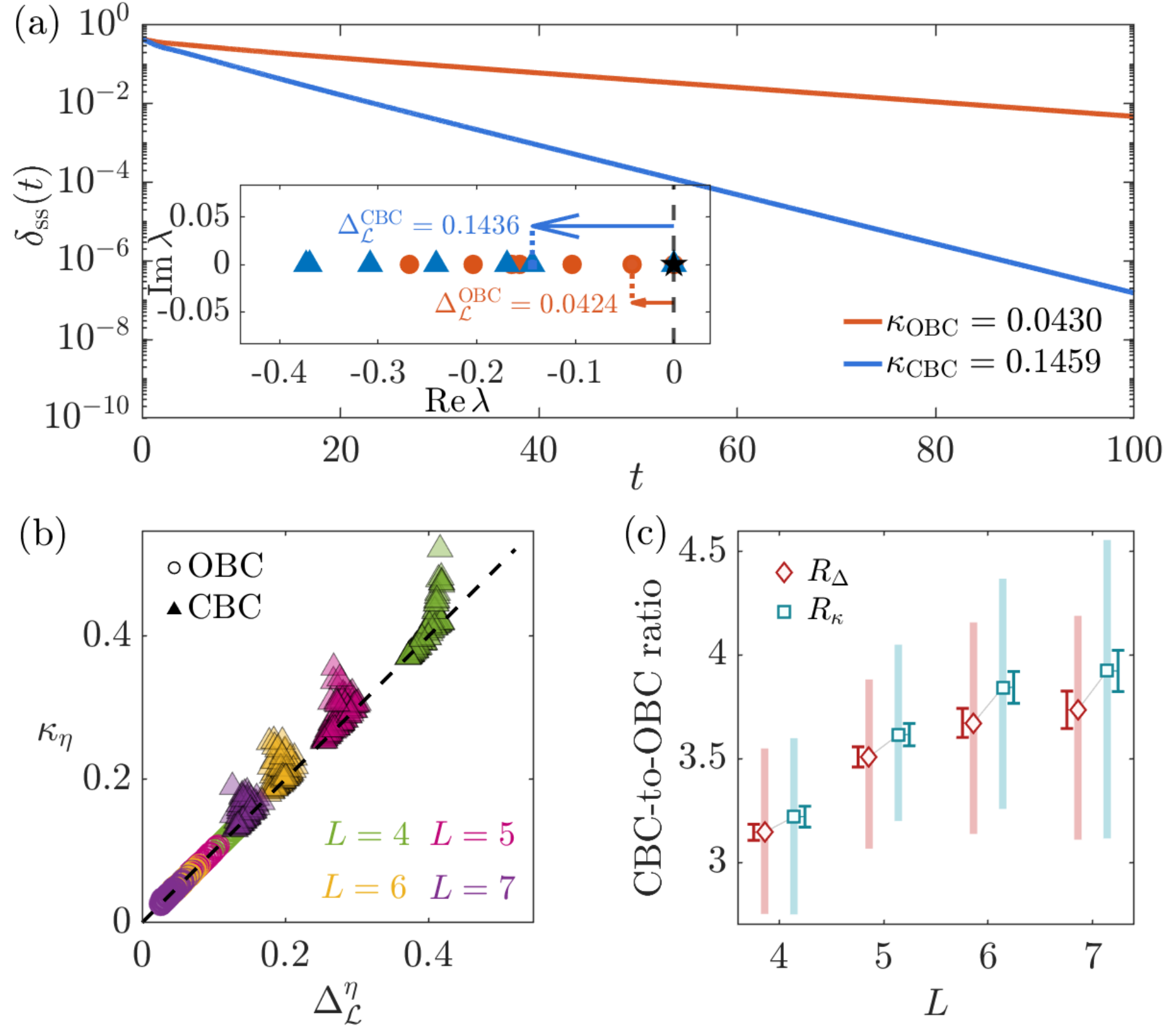}
\caption{
Matter-occupation relaxation rates and Liouvillian gaps under OBC and
reciprocal CBC.
(a) Matter-occupation deviation $\delta_{\mathrm{ss}}^\eta(t)$ for one
$L=7$ realization. The inset shows Liouvillian eigenvalues near
$\lambda=0$, with arrows marking
$\Delta_{\mathcal L}^{\mathrm{OBC}}$ and
$\Delta_{\mathcal L}^{\mathrm{CBC}}$.
(b) Late-time relaxation rates $\kappa_\eta$ versus gaps $\Delta_{\mathcal L}^{\eta}$.
Open circles denote OBC, filled triangles denote CBC, colors distinguish
$L=4,5,6,7$, and the dashed line marks
$\kappa_\eta=\Delta_{\mathcal L}^{\eta}$.
(c) CBC-to-OBC ratios
$R_\Delta=\Delta_{\mathcal L}^{\mathrm{CBC}}/
\Delta_{\mathcal L}^{\mathrm{OBC}}$ and
$R_\kappa=\kappa_{\mathrm{CBC}}/\kappa_{\mathrm{OBC}}$.
Markers, thick bars, and thin error bars denote means, interquartile ranges,
and standard errors, respectively.
All fitted dynamics start from the two leftmost matter sites occupied, all
other matter sites empty, and all link spins in $\ket{\downarrow}$.
Panels (b),(c) use $100$ globally reciprocal Gaussian realizations with exact sample mean zero and sample variance $\sigma_h^2=0.4$ for each
$L$. Other parameters are $J=1$, $\gamma=2$, and $N=2$.
}
\label{fig:relaxation_dynamics}
\end{figure}
Within the fixed-$N$ zero weak-gauge block, relaxation toward the steady state
is governed by the nonzero Liouvillian modes~\cite{Haga2021,Mao2024Liouvillian,Feng2024Boundary}.
For $\eta$ denotes OBC or CBC, let
$\rho_{\mathrm{ss},N}^\eta$ be the fixed-$N$ steady state and let $R_j^\eta$
denote the nonzero right Liouvillian eigenoperators in the same block. Writing
$\rho^\eta(0)=\rho_{\mathrm{ss},N}^\eta+\sum_{j\ge1}c_j^\eta R_j^\eta$ gives
\begin{equation}\label{eq:liouvillian_spectral_expansion}
\rho^\eta(t)
=
\rho_{\mathrm{ss},N}^\eta
+
\sum_{j\ge1}
c_j^\eta e^{\lambda_j^\eta t}R_j^\eta,
\qquad
\mathcal L^\eta[R_j^\eta]=\lambda_j^\eta R_j^\eta .
\end{equation}
Taking $\mathrm{Tr}(\hat n_n\cdot)$ and using
Eq.~\eqref{eq:same_reduced_occupation} gives the deviation from the common
steady occupation,
$N_n^\eta(t)-N_n^{\mathrm{ss}}
=
\sum_{j\ge1}
c_j^\eta e^{\lambda_j^\eta t}
\mathrm{Tr}(\hat n_nR_j^\eta)$.
With Eq.~\eqref{eq:same_reduced_occupation}, the above
expansion converts the
boundary-sensitive nonzero Liouvillian modes into boundary-controlled
matter-occupation relaxation.
We therefore quantify OBC and CBC relaxation by
$\delta_{\mathrm{ss}}^\eta(t)=
L^{-1}\sum_{n=1}^{L}|N_n^\eta(t)-N_n^{\mathrm{ss}}|$.
Starting from the same initial matter occupations,
$\mathrm{Tr}[\rho^\eta(0)\hat n_n]=N_n^{(0)}$, we extract the fitted relaxation
rate $\kappa_\eta$ from the late-time tail
$\delta_{\mathrm{ss}}^\eta(t)\sim A_\eta e^{-\kappa_\eta t}$ and compare it
with the Liouvillian gap
$\Delta_{\mathcal L}^\eta=\min_{\lambda_j^\eta\ne0}
[-\mathrm{Re}\,\lambda_j^\eta]$. For the initial states used below, the
occupation dynamics contains a nonzero contribution from the slowest decaying
mode, so the late-time relaxation rate follows the Liouvillian gap,
$\kappa_\eta\simeq\Delta_{\mathcal L}^\eta$
~\cite{Haga2021,Mao2024Liouvillian,Feng2024Boundary}.
For one representative realization, Fig.~\ref{fig:relaxation_dynamics}(a)
shows this gap-rate correspondence directly. In this realization, $\delta_{\mathrm{ss}}^\eta(t)$ decays for both boundary
conditions toward the common steady occupation given by
Eq.~\eqref{eq:same_reduced_occupation}, and its late-time slopes agree with the
smallest nonzero $-\mathrm{Re}\,\lambda$ in the inset of
Fig.~\ref{fig:relaxation_dynamics}(a). The larger CBC gap therefore yields faster late-time relaxation of the
matter-occupation profile in this realization.

We further verify that both the gap-rate correspondence and the faster CBC
relaxation hold throughout the sampled globally reciprocal realizations. The fitted
rates $\kappa_\eta$ agree with the Liouvillian gaps $\Delta_{\mathcal L}^\eta$ for both boundary conditions, as shown in
Fig.~\ref{fig:relaxation_dynamics}(b). For the same interior realization, CBC gives both the larger gaps and fitted relaxation rates than
OBC, and the sample-averaged CBC-to-OBC ratios remain above unity for
$L=4,5,6,7$, as revealed in Fig.~\ref{fig:relaxation_dynamics}(b),(c). Thus the reciprocal closing
link increases the late-time matter-occupation relaxation rate while leaving the
reduced matter steady occupation unchanged.

In the strong-dissipation limit, the slow matter dynamics reduces to an
effective fixed-$N$ exclusion process whose particle-transfer rates are set by
the link pumping and loss~\cite{Kessler2012generalized,Schutz1997,Derrida1998},
see Supplemental Material~\cite{supplemental}. OBC contains transfers across
the $L-1$ interior links, whereas the reciprocal CBC also includes the closing
transfer between sites $L$ and $1$. For both boundary conditions, the
fixed-$N$ matter-occupation steady distribution remains
Eq.~\eqref{eq:reduced_matter_distribution}, since the
reciprocal closing link leaves the matter-site values $\{X_n\}$ unchanged. With this same steady
distribution, the closing transfer adds a nonnegative term to the Dirichlet
form~\cite{Veerman2020}, see Supplemental Material~\cite{supplemental}. Hence
the CBC exclusion process has a relaxation gap no smaller than
OBC. This strong-dissipation result identifies the
origin of the larger finite-$\gamma$ CBC gap,
$\Delta_{\mathcal L}^{\mathrm{CBC}}>
\Delta_{\mathcal L}^{\mathrm{OBC}}$, observed in the sampled realizations of
Fig.~\ref{fig:relaxation_dynamics}. The boundary closure adds the closing
transfer, while the accumulated field keeps the matter-occupation steady
distribution fixed.

\emph{Conclusions--}
We have investigated a disordered dissipative quantum link chain with locally asymmetric
but globally reciprocal link dissipation. The local dissipative asymmetries
determine a realization-dependent accumulated field, which gives exact
symmetry-resolved steady states and accumulated-field ordered reduced matter
steady occupations through a gauge-generated transformation. The main result
is that the same nonuniform reduced matter steady occupation can be reached
under different boundary conditions, but with distinct relaxation rates toward
it.
The accumulated field fixes the reduced matter steady occupation, whereas the
CBC changes the nonzero Liouvillian spectrum and thereby the
relaxation rate. In the strong-dissipation limit, this boundary-controlled relaxation is
explained by a spectral comparison of effective exclusion processes with open
and cyclic boundaries. The
required matter-link dynamics and link-resolved pumping and loss are accessible
in cold-atom and superconducting quantum-link platforms.

\par\vspace{1\baselineskip}
\emph{Acknowledgments.---}
This work was supported by the Joint Fund of Natural Science Foundation of Shandong Province (Grant
No.ZR2024LLZ004), the National Natural Science Foundation of China(Grant No.12575010, No.12005110),
and the Natural Science Foundation of Shandong Province (Grant No.ZR2020QA078, No.ZR2023MD064).
The Youth Innovation Team Program of Shandong Province (Grant No.2023KJ118)

\par\vspace{1\baselineskip}
\emph{Data Availability.---}
The data that support the findings of this article are openly
available~\cite{dataset_Miao}, embargo periods may apply.

\end{document}